\documentstyle[12pt]{article}

\begin{document}
\setcounter{page}{1} \pagestyle{plain} \vspace{1cm}
\begin{center}
\Large{\bf Is Time Inhomogeneous ? }\\
\small
\vspace{1cm}
{\bf S. Davood Sadatian}\footnote{sd-sadatian@um.ac.ir}\\
\vspace{0.5cm} {\it Department of Physics,
Faculty of Basic Sciences,\\
University of Neyshabur,\\
P. O. Box 91136-899, Neyshabur, Iran
 }\\
\end{center}
\vspace{1.5cm}
\begin{abstract}
In this article, we discuss probability of inhomogeneous time in
high or low energy scale of physics. Consequently, the possibility
was investigated of using theories such as varying speed of light
(VSL) and fractal mathematics to build a framework within which
answers can be found to some of standard cosmological problems and
physics theories on the basis of time non-homogeneity.

\end{abstract}
\newpage

\section{Motivation}
The possibility of breaking physical symmetry in certain positions
in the quantum gravity mechanism has opened a new aspects onto this
realm of physics. This study connect on phenomenological concepts
such as Lorentz invariance violation, in an effort to further
investigate the breaking of principle symmetries in general theory
of relativity. As we know, violation of Lorentz invariance in
high-energy quantum fields theory is predictable [1]. In other
words, Lorentz invariance cannot be a perfect symmetry at all energy
scales. Our main goal in this research is that maybe the effects of
CPT/Lorentz invariance violation can be consideration in nature. In
this case, a new fundamental theory in physics should be favorable
which takes into account the symmetry breaking. On the other hand,
egress of new theories in physics, for example, varying speed of
light theory [5] in physics or fractals mathematic theories [6], has
prompted researchers to study the subject more seriously. Recently
,the possibility of fractal method (non- integer) dimensions is
emphasized. For example, possibility, spatial dimensions are fractal
structures, e.g., volume might be 3D and surface may be considered
to have 2.1D . It should be insisted here that when we considering
fractal dimensions or mass contents, we are concerned ourself only
with very large or very small scales in physics. In other view
point, fractal method in cosmology is a set of minority cosmological
theories which state that the content of matter in the Universe, or
the structure of the universe itself, is a fractal structure. More
generally, A main issue in this field is the fractal dimension of
the Universe or of matter content within it, when study at very
large or very small scales. One A analyze of the SDSS data in 2004
found "The power spectrum is not actually characterized by a single
power law potential but clearly shows curvature" [2]. Their analysis
shows that a fractal structure of galaxies in the universe. In this
regard, theories like Quantum gravity theory[3] are fractal in some
position, for example in the ultra-small near the Planck scale. More
interesting,in July 2008, Scientific American published an article
on Causal dynamical [4] written by the three scientists who proposed
the theory, which consider the universe maybe have a fractal
structure.

\section{Time Non-homogeneity ?}
The main idea in this article is that although the non-homogeneity
of time can be directly obtained from the theory of relativity, we
should instead, according some astronomical observation data, we can
use fractal calculations for cosmological evolution. Moreover,
fractal calculations can also be explain other physical phenomena
such as black holes. Generally, fractal mathematics can be used for
explaining some physical phenomena that cannot be describing by
non-fractal functions or analytical formulas. In following , this
article is divided into four main parts: 1) fractal calculus method,
2) the varying speed of light theory, 3) fractal cosmological
physics and 4) conclusion.

\subsection{Fractal mathematics}
Fractal equations and, in especially, differential fractal relations
have already been performed in many fields of science. For example,
fractal derivatives are used in Classical fluids as good as
dynamical systems. The suitable variation method was used by HE [7]
for differential relations in control method of engineering[7]. The
mathematical structure of fractal calculations has been obtained at
different levels. The well-known case is Riemann-Liouville's
Integration that give as:
\begin{equation}
\frac{d^{-p}f(x)}{dx^{-p}}=\frac{1}{\Gamma(p)}\int_{0}^{x}\frac{f(t)dt}{(x-t)^{1-p}}~,~~~~x>0
\end{equation}
where $p>0$. The Riemann-Liouville's Fractal derivative is obtained
from the following equation:
\begin{equation}
\frac{d^pf(x)}{dx^p}=\frac{d^m}{dx^m}(\frac{d^{-(m-p)}f(x)}{dx^{-(m-p)}})=\frac{1}{\Gamma(m-p)}\frac{d^m}{dx^m}\int_{0}^{x}\frac{f(t)dt}{(x-t)^{1-m+p}}
\end{equation}
here $m$ is an integer number that define  $m-1\leq p< m$. In other
define we have a equation that give as
\begin{equation}
D_{a+,r}^{d}f(r)=(\frac{d}{dt})^m\int_{a}^{r}\frac{f(r')dr'}{\Gamma(m-d(r'))(r-r')^{d(r')-m+1}},
\end{equation}
\begin{equation}
D_{b-,r}^{d}f(r)=(-1)^{m}(\frac{d}{dt})^m\int_{r}^{b}\frac{f(r')dr'}{\Gamma(m-d(r'))(r'-r)^{d(r')-m+1}}
\end{equation}
here $\Gamma(x)$ is Euler's function, and a and b are in interval
$[0,\infty)$. In these definitions, always take, $m = {d} + 1$ ,
that ${d}$ is the integer part of $d$.\\
In the following sections, we consider briefly the varying speed of
light theory, and then use this theory in Section 3.
\subsection{Varying Speed of Light theory}
In recent years, an interesting new theory, Varying Speed of Light
(VSL), has been proposed which assuming varying speed for speed
light($c$). this model can be used for solving the standard
cosmological problems such as Horizon and the cosmological[8].
However, in the most general definition of this theory, it claimed
that within the framework of this theory, laws of physics may be
considered by substituting $c$ (speed of light) in the standard
action with the field $c=c(x^{\mu})$ . As can be seen in this
definition, the speed of light is not constant, but depends on a
specific time and place in space relation( in some models, only the
time dependence is studied for the speed of light). It is more
important that by further taking into account the spatial dependence
of the speed of light, we would open the way for fractal calculus
since in so doing. This means, we actually divide space into
fractals structures by unique properties (for more information, see
[8]). Usually in this theory, the time dependence of the speed of
light is coupled with a parameter of the cosmological scale factor.
For example in some models[5,8] speed of light relation given as
\begin{equation}
c=c_0[{a(t)}]^{n}
\end{equation}
where $a$ is scale factor and $n$ is a constant. We note that $n$ in
above equation determine by observational data fine tuning[8]. In
following, by using this theory and the fractal calculus methods, we
can revisit some of the problems in physics and cosmology. This
means by using fractal mathematics and observation data for
determination of $n$ in equation(5) can solve some physics problems
in fractal differential framework.
\subsection{Fractal Physics and Cosmology}
In this section, we first present a brief description of the fractal
dimensions method for time and space that used in other work.
Consequently, we consider the particular case of this method in
cosmological relations. As an example, we begin with the basic
equations in dynamics of particles, namely, Newton moving's Law. It
can be shown that upon application in fractal calculus. the well-
known form of Newton's relation for gravitational forces is
expressed as [9]:
\begin{equation}
D_{-,s}^{d_{s}(r,s)}D_{+,s}^{d_{s}(r,s)}{\bf
r}(s)=D_{+,r}^{d_{r}}\Phi_{g}({\bf r}(s))
\end{equation}
Here, the previous integer time and space derivatives are replaced
by fractal number that depend on the amount of parameter $d$. In a
generalized case, the Euler equations in classical mechanics
equations can be rewritten in fractal form as:
\begin{equation}
\frac{\partial}{\partial L}(a(L)\frac{\partial d_{s,a}}{\partial
L})+b(L)(L-L_0)d_{a}+c(L)d_{a}^{2}=0
\end{equation}
The above equation can be shown that physical fields may be
reconstructed into their fractal form in which temporal and spatial
dimensions are fractal dimension.\\
In other view point, about impact fractal method and Cosmology, we
can point out a interesting answer to anomalies received in the
ultrahigh energy cosmic rays (UHECR) and TeV-photon spectra[9]. For
example, one has proposed an attractive solution that using with
general relativity and is consistent with the new form of geometry
(Cantorian geometry) of space-time at high-energy scales or
equivalent to fractal methods[9]. There, the author assumed a new
form of dispersion relation for relativistic energy in fractal
calculation as $E^{2\beta}=|p|^{2\alpha}+(m^0_0)^2 $ and obtained an
attractive solution for (UHECR) anomalies in cosmology. Also fractal
formalism offer a strong tool for study the ultraviolet region of
field theory such as: action mechanism, origin of spin method[6,9].\\
In context of fractal cosmology, previous work[10] have been shown
the relation for the relative density of dark contexts. In this
regards, the continuity equation in fractal cosmology model takes
the form as
\begin{equation}
\dot\rho+\left[(L-1)H+\frac{\dot v}{v}\right](\rho+p)=0\,,
\end{equation}
with $v=1$ and $L=4$ (where $L=2+\epsilon$), we get the standard
Friedmann equations in four dimensions.
\section{Summary}
At the start of this article, we asked fundamental questions such as
"Are time and space continuous?", "Why is time non-invertible?", and
"How could time and space dimensions be described?". This research
presents a theory on the nature of time and space based on fractal
geometry formalism. It is obvious that the fractal dimensions method
of space lead to the production of a new define for space-time. In
this discussion, we have assumed that space and time are new-define
fields specified according to the contents of the Universe. Based on
this proposal, the presented model would be a theory to include all
forces within a fractal calculation framework. On the other hand,
the principle lows also holds, e.g., by ignoring the temporal
fractal corrections in formalism, we can obtain the special theory
of relativity. It should be noted that determination of the fractal
parts in fractal calculations is to be adjusted through comparison
with observational data. For example, through the use of varying
speed of light theory, its dependence on space and time fractals. In
this way, many thus far unanswered questions in cosmology can be
addressed through application of fractal physical fields method. The
most significant result of this research is that fractal
calculations formalism maybe used to correct all the theories of
modern physics. The alternative way would be to convert the fractal
equations into integer derivatives without introducing any
corrections.

\end{document}